\documentclass[twocolumn,aps,pra,showpacs]{revtex4}
\usepackage{amsmath}
\usepackage{amssymb}
\usepackage{color}
\usepackage[dvips]{graphicx}

\begin{document}

\title{Spatial entanglement of paired photons generated in cold atomic ensembles}
\author{Clara I. Osorio$^1$, Sergio Barreiro$^1$, Morgan W. Mitchell$^1$ and Juan P. Torres$^{1,2}$}
\affiliation{$^{1}$ICFO-Institut de Ciencies Fotoniques,
Mediterranean Technology Park, Av. del Canal Olímpic s/n, 08860,
Castelldefels, Barcelona, Spain} \affiliation{$^{2}$ Dep. Signal
Theory and Communications, Campus Nord D3, Universitat Politecnica
de Catalunya, 08034, Barcelona, Spain}

\email{clara.ines.osorio@icfo.es}

\begin{abstract}
Cold atomic ensembles can mediate the generation of entanglement
between pairs of photons. Photons with specific directions of
propagation are detected, and the entanglement can reside in any
of the degrees of freedom that describe the whole quantum state of
the photons: polarization, spatial shape or frequency. We show
that the direction of propagation of the generated photons
determines the spatial quantum state of the photons and therefore,
the amount of entanglement generated. When photons generated in
different directions are combined, this spatial distinguishing
information can degrade the quantum purity of the polarization or
frequency entanglement.
\end{abstract}


 \maketitle

\section{Introduction}
The generation of pairs of photons with controllable entanglement
is one of the paramount goals in quantum optics. These states of
light are used as tools to implement new protocols with quantum
enhanced capabilities \cite{bouwmeester1,haroche1}. Although
spontaneous parametric down-conversion (SPDC) is by far the most
widely used source for generating entangled paired photons, in the
last few years, many interesting schemes have been proposed that
make use of Raman transitions on atomic ensembles to generate
entangled pairs of photons.

In these schemes, a classical pump (write) beam impinges on an
ensemble of $N$ atoms, for instance, rubidium or cesium, and it
induces the emission of, at most, a single photon {\em (Stokes
photon)} from one of the atoms \cite{duan1}. Such emission
generates a collective atomic excitation that can be read by a
control beam, which induces the emission of another photon {\em
(Anti-Stokes photon)} entangled with the Stokes Photon. Quantum
correlations mediated by the generation of a collective excitation
in an ensemble of atoms have been observed in polarization
\cite{matsukevich1}, time-frequency \cite{balic1}, and orbital
angular momentum (OAM) \cite{inoue1} degrees of freedom.

In a typical experimental configuration (as shown in Fig.
\ref{raman}), the Stokes and Anti-Stokes photons are detected in a
small section of the full set of directions where the
Stokes/Anti-Stokes photons can be emitted
\cite{duan1,eberly1,scully1}. In most cases, such detection modes
are nearly collinear ($\sim 2$-$3^{\circ}$) with the direction of
propagation of the counter-propagating pump and control beams
\cite{kuzmich1}. But other situations can be considered as well,
as it is the case of transverse emitting configurations, where the
Stokes/anti-Stokes photons propagate transversally to the
pump/control beams \cite{balic1}.

The question arises if the specific non-collinear configuration
used might affect the amount and nature of the generated
entanglement. This is especially important when Stokes/Anti-Stokes
photons selected at different angles are combined to generate new
types of entangled states in a multidimensional Hilbert space
\cite{chen1}. Paired photons generated in different directions
might show {\em azimuthal distinguishing information}, i.e.,
different spatial quantum correlations and amount of entanglement.
This spatial distinguishing information, for instance, can
severely degrade the quality of polarization entanglement, since
the full quantum state that describes the entangled photons is a
nonseparable mixture of polarization and spatial variables. Here
we show that this is the case for highly non-collinear
configurations. The direction of propagation of the pump/control
beams determine a preferred direction, so configurations where the
Stokes and anti-Stokes photons are selected at different angles
with respect to this direction might show different quantum
properties. This is reminiscent of what it happens in
non-collinear SPDC configurations \cite{barbosa1,molina1,osorio1}.

\section{The quantum state of the Stokes/antiStokes pair of photons}
\begin{figure}[t]
\centering
\centering\includegraphics[width=0.95\columnwidth]{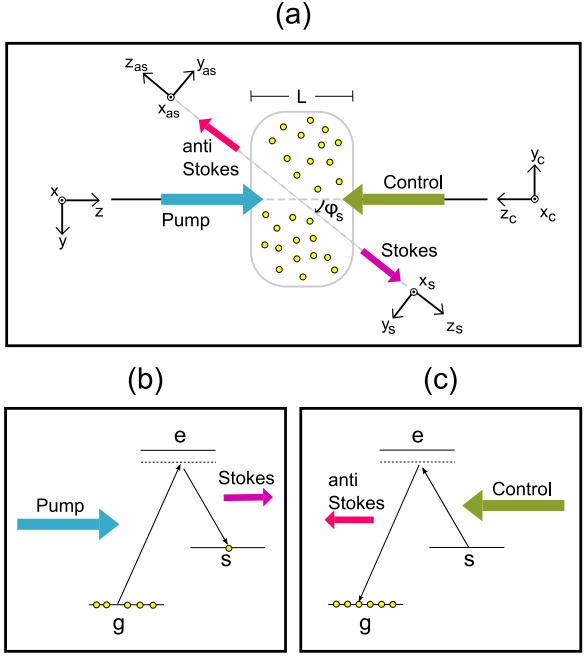}
\caption{General configuration. (a) Sketch of the the geometric
configuration. (b) and (c) Level structure of the atoms that form
the atomic cloud} \label{raman}
\end{figure}

The basic setup considered is shown in Fig. \ref{raman}. An
ensemble of $N$ identical cold atoms is trapped in a
magneto-optical trap (MOT). The atoms have a $\Lambda-$type level
configuration, with two hyperfine ground states, $|g\rangle$ and
$|s\rangle$, and one excited state $|e\rangle$. All atoms are
initially in the ground state $|g\rangle$. Two counter-propagating
CW classical light beams are used to induce the emission of pairs
of photons, the Stokes (s) and anti-Stokes (as) photons.

A weak gaussian pump beam, whose shape in the transverse
wavenumber ${\bf q}=\left( q_x,q_y \right)$ domain is $E_{p}
\left( q_x,q_y \right)$, propagates in the $\hat{z}$ direction.
The action of the pump beam, which is far detuned from the
$|g\rangle \rightarrow |e\rangle$ transition, results in a small
probability of exciting one atom of the cloud and generating by
spontaneous Raman scattering one Stokes photon propagating in the
direction $\hat{z}_s$, which forms an angle $\varphi_s=\varphi$
with respect to the $\hat{z}$ direction. The control beam ${\cal
E}_c$, which is also far detuned from the $|s\rangle \rightarrow
|e\rangle$ transition, propagates in the $-\hat{z}$ direction,
resulting in the generation of an anti-Stokes photon propagating
in the direction $\hat{z}_{as}$, which forms an angle
$\varphi_{as}=\pi-\varphi$ with respect to the $\hat{z}$
direction.

Energy conservation implies,
$\omega_p^0+\omega_c^0=\omega_s^0+\omega_{as}^0$, and the phase
matching conditions impose $k_p^0-k_c^0=k_s^0 \cos
\varphi_s-k_{as}^0 \cos \varphi_{as}$ and $k_s^0 \sin
\varphi_s=k_{as}^0 \sin \varphi_{as}$, where $\omega_i^0$
($i=p,c,s,as$) are the central angular frequencies, and $k_i^0$
are the corresponding wavenumbers at the central frequencies. We
now consider D2 hyperfine transitions in $^{87}$Rb. If the pump
and control beams have the same frequency \cite{balic1}, then the
phase matching conditions allows any angle of emission
$\varphi_{as}=\pi-\varphi$ (always that it is not forbidden by the
transition matrix elements), since one then has $\omega_s^0 \simeq
\omega_{as}^0$ and $k_s^0 \simeq k_{as}^0$.

In order to describe the quantum state of the Stokes photon, we
make use of more convenient transverse wavevector coordinates
 ${\mathbf q}_s=\left(q_s^x,q_s^y \right)$ \cite{torres1}, which
are defined as $q_x=q_s^x$ and  $q_y= q_s^y \,\cos
\varphi_s-k_s\,\sin \varphi_s$, with $k_{s}$ being the
longitudinal wavenumber of the Stokes photon. Similarly for the
anti-Stokes photon, which propagates in the direction ${\hat
z}_{as}$ with longitudinal wavevector $k_{as}$, and transverse
wavevector ${\mathbf q}_{as}=\left( q_{as}^x,q_{as}^y \right)$.

Generation of Stokes/anti-Stokes photons can be accurately
described by means of a) two coupled equations in the slowly
varying envelope approximation for the Stokes and anti-Stokes
electric fields \cite{braje1}, or alternatively, b) making use of
an effective Hamiltonian of interaction and first order
perturbation theory \cite{rubin1}. In this paper, we choose this
second option.

If we assume coherent states for the control and pump beams, with
coherent-state amplitudes $\mathcal{E}_c$ and $ \mathcal{E}_p$
respectively, the effective Hamiltonian in the interaction
picture, that describes the photon-atom interaction, can be
written as
\begin{equation}
H_{I}= \epsilon_0 \int dV \chi^{(3)} \mathcal{E}_{as}^{-}
\mathcal{E}_s^{-} \mathcal{E}_c \mathcal{E}_p + h.c.
\end{equation}
where the electric field operator is written
\begin{eqnarray}
& & {\mathcal E}_{s}^{+} \left( {\bf x}_s, z_s, t \right)= \int
d\omega_s d {\mathbf q}_s a_s \left( \omega_s,{\mathbf q}_s
\right) \nonumber \\
& & \times \exp \left\{ i k_{s} z_s +i {\mathbf q}_s \cdot {\bf
x}_s -i \omega_s t\right\}
\end{eqnarray}
$a_s$ is the annihilation operator of a Stokes photon with
frequency $\omega_s$ and transverse wavenumber ${\bf q}_s$.
$\mathbf{x}_s=(x_s,y_s)$ is the transverse position vector for the
Stokes photon. A similar expression can be written for the
electric field operator ${\mathcal E}_{as}^{+}$, with frequency
$\omega_{as}$ and transverse wavenumber ${\mathbf q}_{as}$.

For non-resonant pump and control beams, the effective
nonlinearity $\chi^{(3)}$ does not depend on the intensity of the
control and pump beams \cite{braje1}. The distribution of atoms in
the cloud is assumed to be gaussian, so the effective nonlinearity
$\chi^{(3)}$ can be written as
\begin{equation}
\chi^{(3)} \left(x,y, z \right) \propto exp \left[
-\frac{x^2+y^2}{R^2}-\frac{z^2}{L^2}\right]
\end{equation}
where $R$ is the size of the cloud of atoms in the transverse
plane $\left(x,y\right)$ and $L$ is the size in the longitudinal
direction.

The spatial quantum state of the generated pair of photons, at
first order of perturbation theory, is $|\Psi\rangle=\int
d{\mathbf q}_s d{\mathbf q}_{as} \,\Phi \left( {\mathbf
q}_s,{\mathbf q}_{as} \right) |{\mathbf q}_s \rangle_s |{\mathbf
q}_{as} \rangle_s$, where the mode function $\Phi$ of the two
photon state is written
\begin{eqnarray}\label{two-photonmode1}
&&\Phi \left( \omega_s,\omega_{as},\mathbf{q}_{s},\mathbf{q}_{as}
\right)= \int d {\mathbf q}_p {\mathbf q}_c  E_{p} \left( {\mathbf
q}_p\right)E_{c} \left( {\mathbf q}_c\right)\\ \nonumber
&&\times\exp{\left(-\Delta_0^2 R^2/4-\Delta_1^2 R^2/4 -\Delta_2^2
L^2/4\right)}
\end{eqnarray}
where
\begin{eqnarray}\label{two-photonmode2}
& & \Delta_0=q_s^x+q_{as}^x \nonumber \\
& & \Delta_1=\left( k_s-k_{as} \right) \sin\varphi+\left(
q_s^y -q_{as}^y \right) \cos\varphi \\
& & \Delta_2=k_p-k_c-\left( k_s+k_{as} \right) \cos\varphi+
\left(q_s^y-q_{as}^y \right) \sin\varphi \nonumber
\end{eqnarray}
and the longitudinal wavevector of the pump beam is written
$k_p=\left[ \left(\omega_p n_p/c\right)^2 -\Delta_0^2-\Delta_1^2
\right]^{1/2}$, $n_p$ is the refractive index at the pump beam
wavelength, and $c$ is the velocity of light in vacuum. Due to the
narrow bandwidth ($\sim$ GHz) of the generated Stokes and
anti/Stokes photons \cite{balic1}, the spatial shape of the mode
function $\Phi$ can be analyzed making the substitution
$\omega_s=\omega_s^0$ and $\omega_{as}=\omega_{as}^0$.

For the sake of clarity, let us consider that the control and pump
beams are gaussian beams with the same beam waist $w_0$ (at the
center of the cloud $z=0$). Gaussian spatial filters with width
$w_1$ describe the effect of the unavoidable spatial filtering
produced by the specific optical detection system used, so that
$1/\left( k_s^0 w_1\right)$ can be considered as the angular
acceptance of the single photon detection system. In most
experimental configurations, $w_1 \sim 50$-$150 \mu$m and the
length of the cloud is a few millimeters long or less. The pump
beam waist is typically $200$-$500 \mu$m. Therefore, the Rayleigh
range of the pump, Stokes and anti-Stokes modes, $L_p=\pi
w_0^2/\lambda_p$ and $L_{s,as}=\pi w_1^2/\lambda_{s,as}$ fulfill
$L \ll L_p,L_{s,as}$. We can thus neglect the transverse
wavenumber dependence of all longitudinal wavevectors in Eqs.
(\ref{two-photonmode1}) and (\ref{two-photonmode2}).

Under these conditions, the mode function $\Phi$ can be written as
\begin{eqnarray}\label{two-photonmode1new}
& & \Phi \left( \mathbf{q}_{s},\mathbf{q}_{as} \right)=
\frac{\left(ABCD \right)^{1/4}}{\pi} \nonumber
\\
& &   \times \exp \left\{ -\frac{A}{4} \left( q_s^x+q_{as}^x
\right)^2-\frac{B}{4}\left( q_s^x-q_{as}^x \right)^2 \right\} \nonumber \\
& & \times \exp \left\{ -\frac{C}{4} \left( q_s^y-q_{as}^y
\right)^2 -\frac{D}{4} \left( q_s^y+q_{as}^y \right)^2 \right\}
\end{eqnarray}
where
\begin{eqnarray}
& & A=\frac{w_0^2 R^2}{2R^2+w_0^2}+\frac{w_1^2}{2} \nonumber \\
& & B=w_1^2/2 \nonumber \\
& & C=w_1^2/2 \nonumber \\
& & D=\frac{w_0^2 R^2 \cos^2 \varphi}{2R^2+w_0^2}+ L^2 \sin^2
\varphi+\frac{w_1^2}{2}
\end{eqnarray}
The state is normalized, i.e., $\int d{\mathbf q}_s d{\mathbf
q}_{as} |\Phi \left( \mathbf{q}_{s},\mathbf{q}_{as} \right)|^2=1$.

Eq. (\ref{two-photonmode1new}) describe the spatial quantum state
of the Stokes/anti-Stokes pair. {\em The important point to be
considered here is that the specific characteristics of the state
depend on the angle of emission}.  For a nearly collinear
configuration ($\varphi=1-3^{\circ}$), which is the case in most
experimental configurations \cite{matsukevich1,kuzmich1,chen1},
the spatial shape of the quantum state shows cylindrical symmetry
in the transverse planes ($x_s,y_s$) and ($x_{as},y_{as}$), since
$A=D$ in Eq. (\ref{two-photonmode1new}). This is not generally
true for other configurations.

\section{Orbital angular momentum correlations}
The azimuthal variation of the spatial correlations translate into
different orbital angular momentum (OAM) correlations between the
Stokes and anti-Stokes photons. In order to make it clearer, let
us consider the case where the anti-Stokes photon is projected
into a gaussian mode $U_g$ with beam width at the center of the
cloud $w_{g}$. This can be achieved by detecting the anti-Stokes
photon with a single-mode fiber, placed after a conveniently
designed optical system. The resulting mode function of the Stokes
photon is written $\Phi_{s} \left( \mathbf{q}_s \right)=\int d
\mathbf{q}_{as} \Phi\left( \mathbf{q}_s,\mathbf{\bf q}_{as}
\right) U_g^{*} \left( \mathbf{q}_{as}\right)$, then
\begin{equation}
\Phi_s \left( \mathbf{q}_{s} \right)= \frac{\left(FG
\right)^{(1/4)}}{\left( 2 \pi\right)^{1/2}} \exp \left[
-\frac{F}{4} \left( q_s^x \right)^2 -\frac{G}{4}
\left(q_s^y\right)^2 \right] \label{stokes}
\end{equation}
with
\begin{eqnarray}
& & F=\frac{4 A B+\left( A+B \right) w_g^2}{A+B+w_g^2} \nonumber \\
 & & G=\frac{4 C D+\left( C+D \right) w_g^2}{C+D+w_g^2}
\end{eqnarray}

\begin{figure}[t]
\centering\includegraphics[width=0.70\columnwidth]{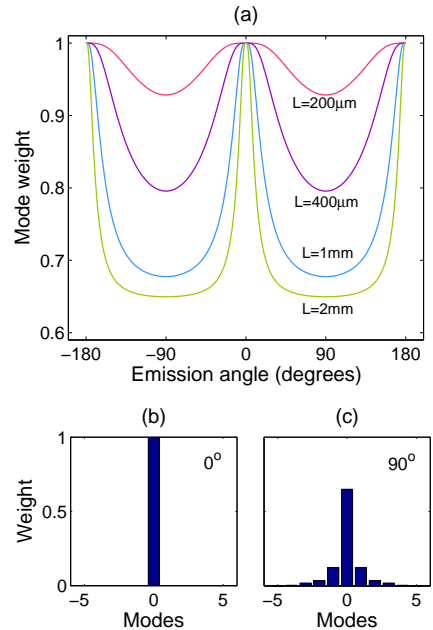}
\caption{a) Weight of the OAM mode $m_s=0$ as a function of the
emission angle, for different values of the length of the cloud of
atoms ($L$). (b) and (c) Two typical OAM distributions of the
Stokes Photon for $\varphi=0^o$ and $\varphi=90^o$, respectively.
In both cases $L=2$ mm. Data: $w_0=100 \mu$m, $R=400\mu$m,
$w_1=100\mu$m and $w_g=500\mu$m.} \label{emisionangle}
\end{figure}

The mode function of the Stokes photon as given by Eq. (\ref
{stokes}) can be described by a superposition of OAM modes
\cite{management} $\Phi_{s} \left( \rho_s,\varphi_s \right)=\left(
2\pi \right)^{-1/2} \sum_{m_s} a_{m_s} \left( \rho_s \right) \exp
\left(i m_{s} \varphi_s \right)$, where ($\rho_s,\varphi_s$) are
cylindrical coordinates in the transverse wavenumber domain, and
$m_s$ is the index of each mode. The weight of each spiral mode
$P_{2m_s}=|a_{2m_s}|^2$, can be found to be
\begin{eqnarray}
& & P_{2m_{s}}=\left( F G \right)^{1/2} \int \rho_s d\rho_s \exp
\left\{ -\frac{F+G}{4} \rho_s^2 \right\} \nonumber
\\
& &  I_m^2 \left[ \frac{G-F}{8} \rho_s^2 \right]
\end{eqnarray}
$I_m$ is the Bessel function of the second kind.

Fig. \ref{emisionangle}(a) shows the weight of the mode $m_{s}=0$
as a function of the angle of emission for different values of the
length of the cloud of atoms. Figs. \ref{emisionangle}(b) and
\ref{emisionangle}(c) shows two typical OAM decompositions. The
pump and control beams are gaussian beams, with $m_c=m_p=0$, where
$m_i$ describes an OAM of $m_i \hbar$ per photon of the
corresponding beam. For nearly collinear configuration, Fig.
\ref{emisionangle} shows that we obtain the expected relationship
$m_s=m_{as}$, while this might not the case for highly
non-collinear configurations.

The results of Fig. \ref{emisionangle} can be understood if we
consider that the effective volume of the interaction is
determined by the effective size of the atomic cloud in the
longitudinal ($L$) and transverse planes ($R$), and by the beam
waist ($w_0$) of the pump beam and control beams, and the size
$w_1$ of the modes of the stokes and antiStokes photons. The
spatial mode function shows cylindrical symmetry when $A=D$ in Eq.
(\ref{two-photonmode1new}), so that the condition
\begin{equation}
L=\frac{R}{\left( 1+2 R^2/w_0^2 \right)^{1/2}} \label{condition}
\end{equation}
if fulfilled. If the beam waist is much larger than the size of
the cloud in the transverse dimensions, the condition turns out to
be $L=R$.

Under the conditions of Eq. (\ref{condition}), the spatial mode
function shows cylindrical symmetry, as for a nearly collinear
configuration, but now, {\em this is true for all possible
directions of emission of the paired photons}. Pairs of photons
emitted in different directions show the same spatial quantum
properties. Any deviation from a spherical-like volume of
interaction, introduces ellipticity in the mode function, and
therefore, azimuthal spatial distinguishing information of pairs
of photons emitted in different directions. In \cite{inoue1} the
relationship $m_s=m_{as}$ is measured. They use a nearly collinear
configuration that, as demonstrated here, it should fulfill this
relationship. Notwithstanding, we predict that this should not be
the case for other non-collinear configurations, as the one used
in \cite{balic1}, if the volume of interaction is highly
elliptical.

\begin{figure}[t]
\centering\includegraphics[width=0.50\columnwidth]{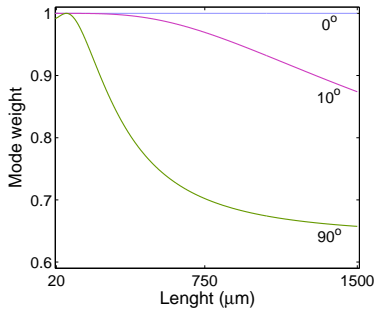}
\caption{a) Weight of the OAM mode $m_s=0$ as a function of the
length of the cloud, for different values of the emission angle
($\varphi$). Data: $w_0=100 \mu$m, $R=400\mu$m, $w_1=100\mu$m and
$w_g=500\mu$m.} \label{length}
\end{figure}

Fig. \ref{length} shows the weight of the mode $m_{s}=0$ as a
function of the length of the cloud for different values of the
angle of emission. For a collinear configuration, $m_s=m_{as}=0$
for any value of the length of the cloud. When the angle of
emission increases, the OAM distribution now depends on the length
of the cloud, but the change of the length of the cloud affects
very weakly the mode weight when the length of the cloud is much
longer than the relevant parameters: $w_1$, $w_g$ and $R$. This is
specially evident for the case of the angle of emission
$\varphi=90^0$.

\section{Spatial entanglement}
\begin{figure}[t]
\centering\includegraphics[width=1\columnwidth]{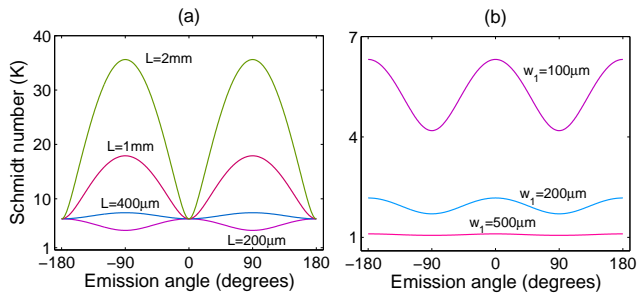}
\caption{Amount of spatial entanglement (Schmidt Number, K) of the
Stokes/anti-Stokes pair for different angles of emission
$\varphi$. a) For various values of the length of the cloud (as
indicated by the label).  $w_1=100 \mu$m. b) For various values of
the pump beam width (as indicated by the label). Length of the
cloud: $L=200 \mu$m. In both cases: $R=1000 \mu$m and $w_0=500
\mu$m.} \label{schmidtN}
\end{figure}

Let us consider in more detail how it changes the quantum state of
different pair of photons emitted along different directions of
propagation. The amount of spatial entanglement embedded in the
quantum state can be quantified by the Schmidt number
\cite{schmidt} $K=1/\sum_n \lambda_n^2$, where the eigenvalues
$\lambda_n$ comes from the decomposition $\Phi \left(
\mathbf{q}_{s},\mathbf{q}_{as} \right)=\sum_n \sqrt{\lambda_n} f_n
\left( \mathbf{q}_{s} \right) g_n \left( \mathbf{q}_{as} \right)$.
Taking into account Eq. (\ref{two-photonmode1new}), one can find
that \cite{woerdman1,chan1}
\begin{equation}
K=\frac{\left(A+B\right) \left( C+D \right)}{4 \left(
ABCD\right)^{1/2}}
\end{equation}
Fig. \ref{schmidtN} shows the value of the Schmidt number as a
function of the angle of emission $\varphi$ for different values
of the length of the atomic cloud and the filter width.
Importantly, the degree of spatial entanglement of the two-photon
state shows azimuthal variations, depending on the direction of
emission of the Stokes/anti-Stokes photons. For $L < R \left(
1+2R^2/w_0^2 \right)^{-1/2}$, as dictated by Eq.
(\ref{condition}), the maximum amount of entanglement is achieved
for nearly collinear configurations ($\varphi \sim 0^{\circ}$). In
other words, the Schmidt decomposition contains more modes in a
nearly collinear configuration, than in a transverse emitting
configuration ($\varphi=90^{\circ}$), showing therefore a higher
degree of entanglement. If we increase the length of the cloud,
the amount of entanglement decreases, as well as the azimuthal
variability of the entanglement. If Eq. (\ref{condition}) is
fulfilled, the amount of spatial entanglement is constant for all
angles of emission, For $L > R \left( 1+2R^2/w_0^2
\right)^{-1/2}$, the maximum amount of entanglement is achieved
for transverse emitting configurations. Notice also that strong
spatial filtering, i.e. increasing $w_1$, also diminishes both the
amount of entanglement and its azimuthal variability.

\section{Conclusions}
We have shown that pairs of entangled Stokes/anti-Stokes photons,
when generated in different directions of propagation, show
different quantum spatial properties due to the presence of
ellipticity of the mode function, i.e. {\em azimuthal spatial
distinguishability}. The degree of ellipticity, and azimuthal
distinguishability, depends on the shape of the volume of
interaction of the atom-light interactions. It is negligible for
nearly collinear configurations, and for highly noncollinear
configurations in spherical-like clouds of atoms.

The measurement of the ellipticity of the mode function, which
would translate in the observation of paired photons with OAM
correlations that do not fulfill the relationship $m_s=m_{as}$ is
within the availability of current experimental configurations for
highly non-collinear configurations, such as transverse emitting
configuration\cite{balic1}, when highly elliptical atom clouds are
considered. In SPDC configurations, many experimental
configurations make use of highly elliptical volumes of
interaction. In this regime, the ellipticity of the mode function
has already been experimentally verified \cite{barbosa1,molina1}.

This effect might have an important impact on the application of
quantum information protocols that make use of orbital angular
momentum correlations \cite{molina2}. The azimuthal distinguishing
information introduced by the direction of emission can affect the
quantum properties of polarization-entangled photons when these
photons are generated with different angles of emission, as it is
the case in \cite{chen1}.  If the volume of interaction is
spherical-like ($R \simeq L$), the realization of high-dimensional
entanglement by selecting several spatial modes (directions of
emission), as proposed in \cite{chen1}, can be achieved without
introducing spatial distinguishing information between different
pairs of photons, which can degrade the quality of the
entanglement generated. On the other hand, the presence of
ellipticity of the mode function as a function of the emission
angle, could restrict the angles of emission accessible for
generating a polarization entangled state with a degree of
concurrence above a certain prescribed level.

\section{Acknowledgements}
This work was supported by projects FIS2007-60179, FIS2005-03394
and Consolider-Ingenio 2010 Quantum Optical Information
Technologies (QOIT) from the Government of Spain, and by the
European Commission (Qubit Applications (QAP), Contract No.
015848). Sergio Barreiro is on leave from the Universidad de
Montevideo.

\end{document}